# EXPERIMENTAL PLAN TO VERIFY THE YPCP MODEL: "YUKAWA PICO CHEMISTRY and PHYSICS" IMPLICATIONS IN THE CF-LENR FIELD.

Jacques Dufour[1], Xavier Dufour, Denis Murat, Jacques Foos
*CNAM - Laboratoire des Sciences Nucléaires - 2 rue Conté - CC304A - 75003 Paris*

*Abstract: In the CF-LERN field (Cold Fusion and Low Energy Nuclear reactions) many experimental results are available: unexplained energy production, presence of unusual patterns of classical fusion reaction products, isotopic composition variations, sporadic emission of nuclear radiations ($\alpha$, $\beta$ and $\gamma$)... These effects are not always observed, for similar experimental conditions. Should a fundamental reason exist for these effects to occur, funding would be justified, to make them repeatable and more intense (this step being likely to be a trial and error process, might require a substantial amount of money). In this article, a possible fundamental explanation of the phenomenon is described, together with the experimental plan to assess it.*

## Introduction:

One of the main conceptual problem to be addressed (and solved) in the field of CF-LENR, is: "how can huge potential barriers be overcome" (some 0.3 MeV in the case of d/d fusion and 30 MeV in the case of d/Pd nuclear reactions). Part of the answer is probably to be found in specificities of deuterons behavior in a lattice (deuteron/phonon interaction, resonant electromagnetic-dynamics…), that could increase the otherwise very low probability of reaction. If an attractive (and yet undiscovered) potential exists, with a pico-meter range and a coupling constant in the order of magnitude of the EM coupling constant, these probabilities could be considerably increased and result in macroscopic effects, potentially useful for technical applications. The experimental approach described below, aims at unveiling and characterizing such a potential. When positive, the path would be opened to pico chemistry and physics (YPCP).

## Theoretical part:

### Basis to assess the effects of the new potential :

A first attempt to describe the expected effects of such a potential had been presented [1]. Recently [2], arguments have been reported, (in the frame of SME -Lorentz symmetry violation) for the possible existence of a Yukawa type of potential, with a range extending farther than fm and up to pm distances. The YPCP model has been built on this possibility. It aims at proposing and designing unambiguous experiments, to assess the effects and hence the reality of this potential.

---

[1] corresponding author - phone : (+33) 1 40 27 29 15 , mail : dufourj@cnam.fr  or  3jdufour@orange.fr

From [2], the following Yukawa type of potential, acting between nucleons, was extracted (and is called along this article: weak long range Yukawa potential):

$$V_{WLY} = -Cg^2 \frac{e^{-\nu r}}{r}$$   (1)   (C being a constant, $g^2$ the coupling constant of the strong nuclear force and $\nu$ ($\geq 0$) the reverse of the range $\rho$'.

### *The model used to test the YPCP hypothesis:*

A proton **p** (or a deuteron **d**) enters the electronic system of an atom $\mathcal{A}$ (atomic number **Z**, mass number **A** and radius **a**). At distance **r**, the charge repartition of the electrons of $\mathcal{A}$ is supposed to be uniform (*figure 1*), resulting in a simple way to evaluate their screening effect on the repulsive potential of the nucleus of $\mathcal{A}$. The total energy of the proton is the sum of the attractive nuclear and weak long range Yukawa potentials (acting on nucleons) energies ($E_N$ and $E_{WLY}$), the repulsive Coulomb potential (acting on charges) energy ($E_C$), and the repulsive Kinetic energy of **p/d** ($E_K$) evaluated from Heisenberg uncertainty principle. **$m_p$** being the proton mass, $\kappa$ a constant (1 for a proton entering $\mathcal{A}$ and 2 for a deuteron), the total energy $E$ of the proton (deuteron) is:

$$E = E_{NY} + E_{WLY} + E_C + E_K = -\kappa g^2 \frac{e^{-\mu r}}{r} - Cg^2 \kappa A \frac{e^{-\nu r}}{r} + q^2 Z \left(1 - \left(\frac{r}{a}\right)^3\right)\frac{1}{r} + \frac{\hbar^2}{2\kappa m_p r^2}$$   (2)

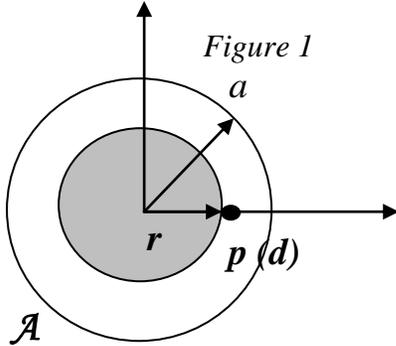

*Figure 1*

The range of the nuclear potential being very small, the impacting proton (deuteron) is supposed to act only with the nucleon of $\mathcal{A}$ at the contact. $q^2$ is the electro-magnetic coupling constant and $\nu$ the reverse of the range of the weak long range Yukawa potential with $\nu > 0$ (unlimited range). The total energy $E$ has been calculated as a function of **r** for different values of **C** and $\nu$. As the weak long range Yukawa potential has not been seen outside atoms, the following constraint has been taken:   $C \leq 8.144 * 10^{-8} * e^{70\nu}$   (energy less than 30 meV, between the 2 tritium atoms of a tritium molecule, $\frac{1}{\nu}$ being expressed in pm).

### *Main experimental results expected from the model:*

Two cases were considered, representative of the experimental situations to be examined:

The deuterium/palladium case (gas loading experiments YPC: Yukawa Pico Chemistry).
The deuterium/deuterium case (proton beam experiments YPP: Yukawa Pico Physics).

For these 2 calculations, following values have been used: Range ($\frac{1}{\nu}$) = 1.3 pm and C=1.101*10$^{-3}$, corresponding to a coupling constant value of 5.048*10$^{-28}$ (to be compared to the electro-magnetic coupling constant: 2.308*10$^{-28}$). These sets of values could be fine-tuned to fit the experimental data (enthalpy of reaction, taken in that case to be some 9 keV see below).

*Deuterium/palladium case (Figure 2):*

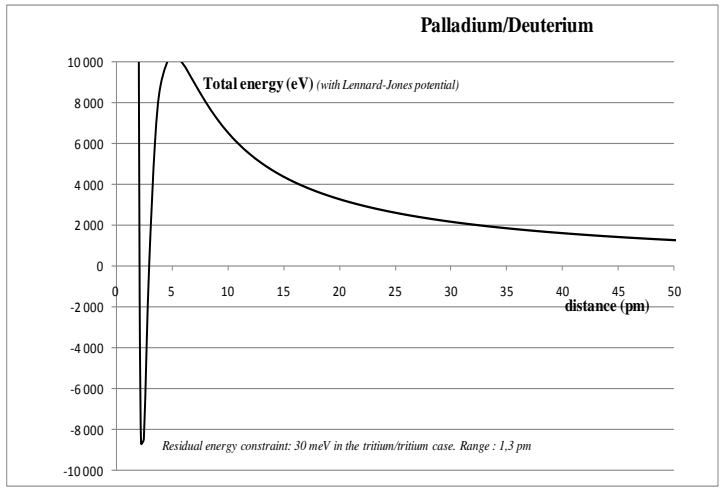

*Figure 2*

It can be seen that the total energy remains positive and increases, from 0 at the periphery of the palladium (179 pm) up to 10000 eV at some 6 pm from the nucleus. The total energy then decreases to negative values and the deuteron could reach the nucleus, resulting in a nuclear reaction. It is thought that this is prevented by the Pauli exclusion principle, the K electrons having few energy levels available. To take this into account, a Lennard-Jones type of potential has been added to the other potentials. A stable bound state ($\left[ {}^{106}_{46}Pd, {}^{2}_{1}H \right]$, reaction enthalpy of some 9000eV) could then be justified.

*Deuterium/deuterium case (Figure 3):*

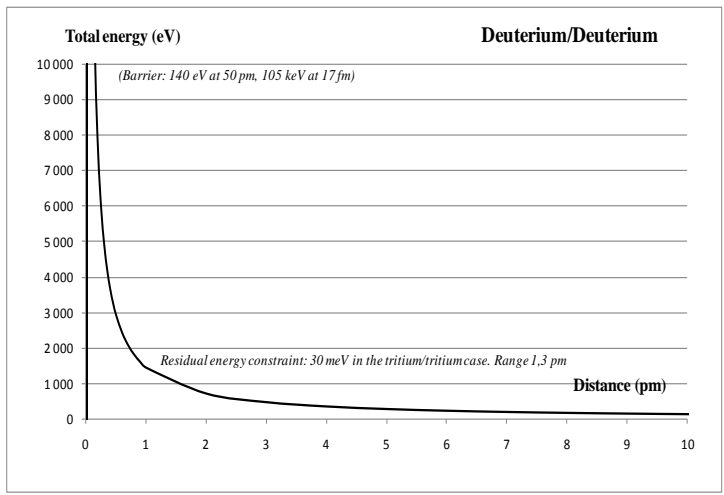

*Figure 3:*

It can be seen that the total energy remains positive and increases from 0 at the periphery of the deuterium (73 pm) up to 140 eV at some 10pm from the nucleus and 113 keV at some 15 fm from it. The total energy then decreases sharply and a d/d fusion reaction can take place. The potential barrier to be overcome is rather low and thin. This could be an explanation of the increase of the fusion reaction cross section at low energy of the deuteron, observed with deuteron beam experiments [3]

## <u>The neutron capture cross sections and the weak long range Yukawa potential:</u>

The weak long range Yukawa potential should have an influence on neutron capture cross sections. The neutron capture cross sections can be viewed as the result of purely geometrical collisions followed by nuclear reactions, increasing the geometrical cross sections. It is however striking to see [4] that for low energies of the neutron, ($10^{-5}$ to $10^{-1}$ eV), there is a considerable increase of the capture cross section $\sigma_{meas.}$. This can be simply interpreted by the existence of the, yet unknown, weak long range Yukawa potential. $\rho' = \frac{1}{v}$ being the range of this potential, $V_N$ the velocity of the neutron and $V_I$ the velocity of the interaction, the geometrical cross section $A_G$ is (*Figure 4*):

$$A_G = \frac{\Pi \rho'^2}{1 + \left(V_N / V_I\right)^2} \qquad \text{(radius } r_G = \frac{\rho'}{\sqrt{1 + \left(V_N / V_I\right)^2}} \text{)}$$

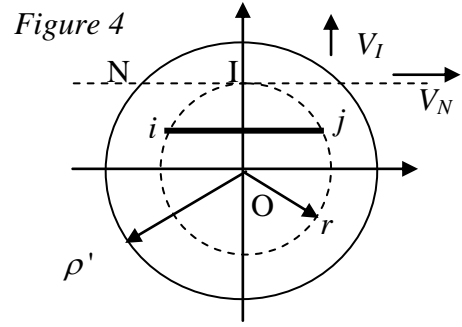

Figure 4

The average interaction time being $T = \frac{ij}{V_N} = \frac{r_G \sqrt{3}}{V_N}$,

the probability of neutron interaction and thus of neutron capture increases when $V_N$ decreases. The true reaction cross section can thus be visualized by normalizing the measured neutron capture cross sections $\sigma_{meas.}$ to a reference velocity of the neutron (chosen for $\left(E_N\right)_{ref} = 10^6\, eV$, with $V_N = \sqrt{\frac{2E_N}{m_n}}$ ).The normalized true reaction cross section $\sigma_{nor.}$ is calculated from the measured neutron capture one $\sigma_{meas.}$, by : $\qquad \sigma_{nor.} = \sigma_{meas.} \dfrac{T\left(E_N\right)_{ref}}{T\left(E_N\right)} \qquad$ (3) $\qquad$ ($V_I$ is taken as 0.1*c, c: speed of light)

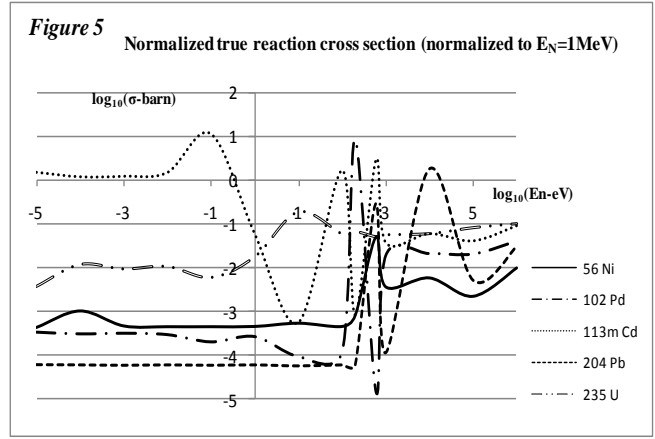

It can be seen (*Figure 5*) that, for each element, $\sigma_{nor.}$ is practically constant for low energies of the neutron (<0.1 eV). In the range 0.1 to $10^4$ eV, resonances are observed (depending upon the element). At higher energies (> $10^5$ eV), $\sigma_{nor.}$ stabilizes at somewhat higher values than for very low energies (with the exception of $^{113m}Cd$). The variations of the neutron residence time, (depending upon its energy), in the range of the weak long range Yukawa potential are a straightforward explanation of this behavior. This could be a first macroscopic and experimental proof of YPCP. Looking then to the microscopic side, a candidate for the required boson carrying the interaction could be creation and exchange of electron-positron pairs (neutral). In the context of Feynman's equation for the Yukawa potential, the effect extends into the picometer range. This would have to be examined in detail when experiments are positive [5].

### *The YPCP conjecture and the CF-LENR field:*

*Palladium gas loading:*

The Coulomb barrier to be overcome, for a reaction between Pd and d to occur, would be some 30 MeV. With the weak long range Yukawa potential, this barrier could be reduced to a few keV (see *Figure 2*). It is then hypothesized that, with this reduced barrier, collective or resonant effects in the lattice, as developed in [6] for instance, can trigger pico-chemical reactions, at the macroscopic level. These reactions are likely to emit X-Rays in the keV level of energy, which could sustain the reaction once initiated. This would be a simple explanation of the phenomenon of "heat after death" already observed [7], [8]. During this episode, where energies (X-Rays) of several keV can interact with the lattice, it is likely that part of the helium4 present in the virgin palladium could be expelled into the gas phase.

The products of these hypothetical YPCP reactions (that are to be characterized as explained later) are probably rather stable: the deuteron would have a high barrier to overcome (Pauli exclusion principle), before reacting with the Palladium nucleus. The probability of overcoming this barrier is nevertheless not zero. The nuclear reaction would be:

$$\left[ {}^{106}_{46}Pd + {}^{2}_{1}H \right] \rightarrow {}^{108}_{47}Ag * \text{ (excess energy* 10.8 MeV)} \qquad (3)$$

${}^{108}_{47}Ag *$ then decaying through $\alpha$ emission (energy 10.8 MeV/helium4 production), ultimately yielding ${}^{104}_{46}Pd$ through $\beta^-$ decay (1.4 MeV, and X-Ray emission) of intermediate ${}^{104}_{45}Rh$. (4) This type of reaction would cause a shift in palladium isotopes.

This could explain observations made in SPAWAR experiments, where entities such as $\left[ {}^{106}_{46}Pd, {}^{2}_{1}H \right]$, formed during electrolysis on the cathode, could pass into the electrolyte and then react according to (3) and (4), in unexpected places in the experiment [9].

Finally, it is thought that the situation in gas loading experiments is also representative of the situation in electrolysis and gas discharges, where energies of the deuteron are also in the order of a few eV.

*Proton (deuteron) beam experiments:*

Typical experiments with deuterons beams, are run with energies of the deuterons of hundreds of keV. Recent experiments [3] have been run with deuterons energy round 5 keV. It has been shown that the reaction cross section is several orders of magnitude higher than was expected from results at higher energies. This has been attributed to a screening potential of the electrons, in the order of hundreds of eV. The action of the weak long range Yukawa potential could add to this screening effect. Lowering the energy of the impacting deuteron, could result in the increase of the measured reaction cross section, by increasing the interaction time with the weak long range Yukawa potential (as for the neutron capture cross sections). Moreover, the full Coulomb barrier (some 300 keV) would be reduced down to some 110 keV. It is then thought that the use of deuterons of energies in the range 15 to 150 eV, could increase the d/d fusion cross section observed in [3].

## *Experimental part:*

### *Heat measurements:*

They will be made using an ice calorimeter built and characterized for that purpose. A detailed description of this device is given in [10]. On the one hand, this calorimeter is perfectly adapted for gas loading experiments and has a high absolute precision. On the other hand, gas loading experiments always give a response, more or less intense (depending on the quality of the palladium black used). This quality (and the thermal response) could be increased by preventing sintering (supported palladium on carbon black [11], or ZrO2 [7]). If possible and to simplify the trace analysis that are going to be performed on the processed samples, pure palladium black (activated at 230°C, under $10^{-1}$ mb pressure) will be used, thus eliminating the pollution by the catalyst support.

Palladium black gas loading will be performed with deuterium and hydrogen [10]. Particular attention will be paid, to the thermal behavior of the system after loading (heat after death [8]). To increase the thermal response of the system, round 20 g of palladium black will be used for each experiment (in that case, an absolute precision on thermal measurements of ± 3% is expected). The optimal palladium quantity to be used will be constrained by its cost.
In parallel experiments (that is with no heat determination) the possible emission of low energy X-Ray during gas loading, will be assessed.

After experiment, the processed samples will be recovered for analysis, allowing the measurement of the enthalpy of reaction. The gas phase will also be recovered in a stainless steel container for helium4 analysis. An overall helium4 balance will be tempted. See below analysis for more details.

### *Charged particles emission:*

The possibility of the emission of charged particles caused by the impact of low/medium energy protons or deuterons (between 15 and 150 eV) on a metallic target will be assessed. Use will be made of a device that has been built to generate protons with energies in that range and could demonstrate the occurrence of true cold d/d fusion reactions.
The metallic target will be either a metallic foil (palladium) or a metallic powder (palladium black, palladium hydride (deuteride), titanium hydride (deuteride).

*Description of the proton (deuteron) generator:*

A concept similar to the Penning ion source has been used: a cloud of protons or deuterons (and not a beam) is generated. The protons or deuterons then impact, with a controlled energy, a target made from the metal to be studied. *Figure 6* is a description of the device.

*Figure 6*

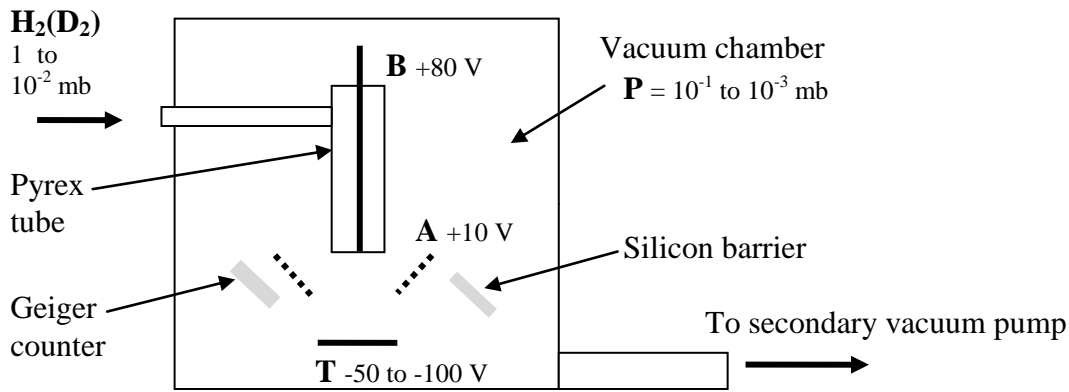

Hydrogen (deuterium) leaking at a pressure between 1 and $10^{-2}$ mb, enters a vertical pyrex tube (12 mm diameter). At the outlet of the tube, the hydrogen expands into a vacuum chamber, maintained at a pressure between $10^{-1}$ and $10^{-2}$ mb by a primary vacuum pump. The use of a secondary vacuum pump allows the pressure in the vacuum chamber to be lowered to values in the range $10^{-2}$ and $10^{-3}$ mb.

The electrode **B** is maintained at a voltage round +150 V. The electrode **A** consists of 4 tungsten filaments, coated with barium oxide and heated (DC +10 to +12 V) at temperatures up to 1000°C. The barium oxide coating allows a copious electronic emission round 1000°C, thus preventing vaporization of the tungsten filaments. The electrode **T** is the target, made from the metal to be studied. It is maintained at a voltage round -150 V. Under these conditions, an electronic current flows from **A** to **B** and ionizing collisions occur with the entering hydrogen (deuterium). The target electrode **T** (metal to be studied) then attracts the protons (deuterons) formed, which hit it with energies of several tens of eV. The distance between **T** and **B** can be varied from 2 to 10 cm.

Preliminary tests have shown that, at a hydrogen pressure round $10^{-1}$mb in the vacuum chamber, the electronic current is round 300 mA and the protonic current round 500 μA. The distance between **A** and **B** was 2 cm and the mean hydrogen free path round 1000μm. The average energy of the protons can thus be estimated to be 15 eV in that case.

A Geiger Muller counter and a silicon barrier detector are installed in the vacuum chamber to detect radiations and charged particles. A helium3 neutron detector is installed outside the vacuum chamber to detect neutron emission. Nuclear radiations detectors and the expertise to use them are available in the laboratory.

After experiment, the processed metal will be recovered for analysis.

### *Description of the analysis to be performed:*

From the conclusions of the YPCP model, it is anticipated that 2 kinds of reactions will occur under the experimental situations that are going to be studied:

-pico-chemical reactions (YPC), resulting from the binding of a proton (deuteron) close to the nucleus of the treated atom (in the order of some pm). This non nuclear reaction would have an enthalpy of reaction of a few keV. X-Rays of that energy could be emitted. Deuterium should be much more reactive than hydrogen (which is likely not to react).

-true nuclear fusion reactions, resulting from pico physics (YPP). These dd fusion reactions would yield the usual reaction products (protons, helium3 and neutrons) with associated energies. Only deuterium is expected to react.

In both experimental situations, the two types of reactions might happen (explaining the great confusion induced by experimental results). However, the experiment using gas loading of the palladium would preferentially yield YPC reactions (as in electrolysis and electrical discharges). On the contrary, in the experiments where energetic protons hit a metallic target, fusion reactions would be the main reaction channel, through YPP.

*Analysis in the case of gas loading:*

In that case, main reactions expected are pico-chemical reactions. These reactions should yield very special atoms. This will be illustrated by considering, as an example, the $^{106}_{46}Pd$ isotope reacting with deuterium:

$$^{106}_{46}Pd + ^{2}_{1}H \rightarrow \left[ ^{106}_{46}Pd, ^{2}_{1}H \right] + a\ few\ keV$$

The entity written $\left[ ^{106}_{46}Pd, ^{2}_{1}H \right]$, represents a "pseudo-silver" atom. Its pseudo nucleus is composed of a $^{106}_{46}Pd$ nucleus, surrounded by a $^{2}_{1}H$ nucleus turning around at very short distance (a few pm). The mass of this "pseudo-silver" atom (#107.917478 that is the mass of $^{106}_{46}Pd$ + the mass of deuterium $^{2}_{1}H$, within a few keV, binding energy of the entity "pseudo-silver"), is higher than that of the corresponding silver atom $^{108}_{47}Ag$ (107.905952). The 47 surrounding electrons would see a positive charge equivalent to 47 protons, which can no longer be considered as a point charge. It is expected that the outermost electronic layers will closely look like those of $^{108}_{47}Ag$. The differences will increase, when going deeper into the electrons levels. The K levels are likely to be strongly altered.

To characterize this "pseudo-silver", its electronic system and mass will be assessed, using:

      ICP-AES or colorimetric measurements that see the outermost electronic levels.
      XRF (X-Ray fluorescence) or Auger that see intermediate electronic levels (L and M)
      ICP-MS high resolution that sees the mass.

    - XRF [12] or Auger would be the most preferred method to assess the electronic system, but are not available in the lab. It is worth noting that Auger has shown a signal, resembling that of silver, on a palladium cathode used in heavy water electrolysis [13].
    - ICP-AES (available in the lab) is highly selective (an extremely small departure from the silver outermost electronic system will cause a big shift of the characteristic silver peaks) and is not usable in that case. Colorimetric measurements (available in the lab) will thus be used, but there are tedious, non selective and not very sensitive.
    - ICP-MS high resolution is not available in the lab, but the mass scans required could be obtained from another lab, on samples showing anomalies in the "chemical" characterization.

The products of the pico-chemical reactions are present in small amounts in the processed sample. The palladium matrix will thus be selectively extracted (DMG).

To evaluate the possible occurrence of helium4 producing reactions in the gas loading case, a helium4 balance will be tempted (helium4 in virgin palladium black and virgin deuterium, helium4 in processed palladium black and processed deuterium). Analyses for helium4 in the palladium black and the deuterium are not available in the lab. They are available at [14].

*Analysis in the case of charged particles emission:*

The processed targets samples will be analysed in the same way as for the gas loading experiments.

## <u>*Conclusion:*</u>

When positive, the proposed experiments would prove the reality of the YPCP working hypothesis and the existence of the weak long range Yukawa potential. Once this necessary scientific confirmation is obtained, funding could be obtained for developments that can be envisaged along 2 paths:

- nuclear wastes remediation [15] and small heat generators (if metals other than palladium can be used). These developments could be achieved through gas loading, electrolysis, electrical discharges …

-true cold fusion d/d reactions (deuterons beam experiments) ultimately leading to a d/d fusion reactor based on well mastered technologies, using usual temperature and pressure conditions.